\documentclass[multphys,vecphys]{svmult}

\usepackage{makeidx}     
\usepackage{graphicx}    
\usepackage{multicol}    

\makeindex             

\begin{document}

\title*{UYSO 1 - An Extremely Young Massive Stellar Object}
\author{J. Forbrich\inst{1}\and K. Schreyer\inst{2} 
\and B. Posselt\inst{3} \and R. Klein\inst{4} \and Th. Henning\inst{5}}
\institute{Max-Planck-Institut f\"ur Radioastronomie, Auf dem H\"ugel 69, D-53113 Bonn, Germany
\texttt{forbrich@mpifr-bonn.mpg.de}
\and Astrophysikalisches Institut und Universit\"ats-Sternwarte, Schillerg\"a{\ss}chen 2-3, D-07745 Jena, Germany
\texttt{martin@astro.uni-jena.de}
\and Astrophysikalisches Institut und Universit\"ats-Sternwarte, Schillerg\"a{\ss}chen 2-3, D-07745 Jena, Germany
\texttt{posselt@astro.uni-jena.de}
\and Max-Planck-Institut f\"ur extraterrestrische Physik, Giessenbachstra{\ss}e, D-85748 Garching, Germany
\texttt{rklein@mpe.mpg.de}
\and Max-Planck-Institut f\"ur Astronomie, K\"onigstuhl 17, D-69117 Heidelberg, Germany
\texttt{henning@mpia.de}}
%
%
\maketitle

In the course of a comprehensive mm/submm survey of massive star-forming
regions, a particularly interesting object has been found in the surroundings of the bright FIR source IRAS 07029-1215, in a distance of 1 kpc. The object -- named UYSO 1 (\textsl{Unidentified Young Stellar Object 1}) -- is cold ($T=40$ K), it has a massive envelope, and it is associated
with an energetic molecular outflow. No infrared point source has been detected at its position for $\lambda \le 20$ $\mu$m. Therefore, it is a very good candidate for a member of the long searched for group
of massive protostars.

\begin{figure}
\centering
 \begin{minipage}[t]{9cm}
     \resizebox{10cm}{!}{\rotatebox{-90}{\includegraphics{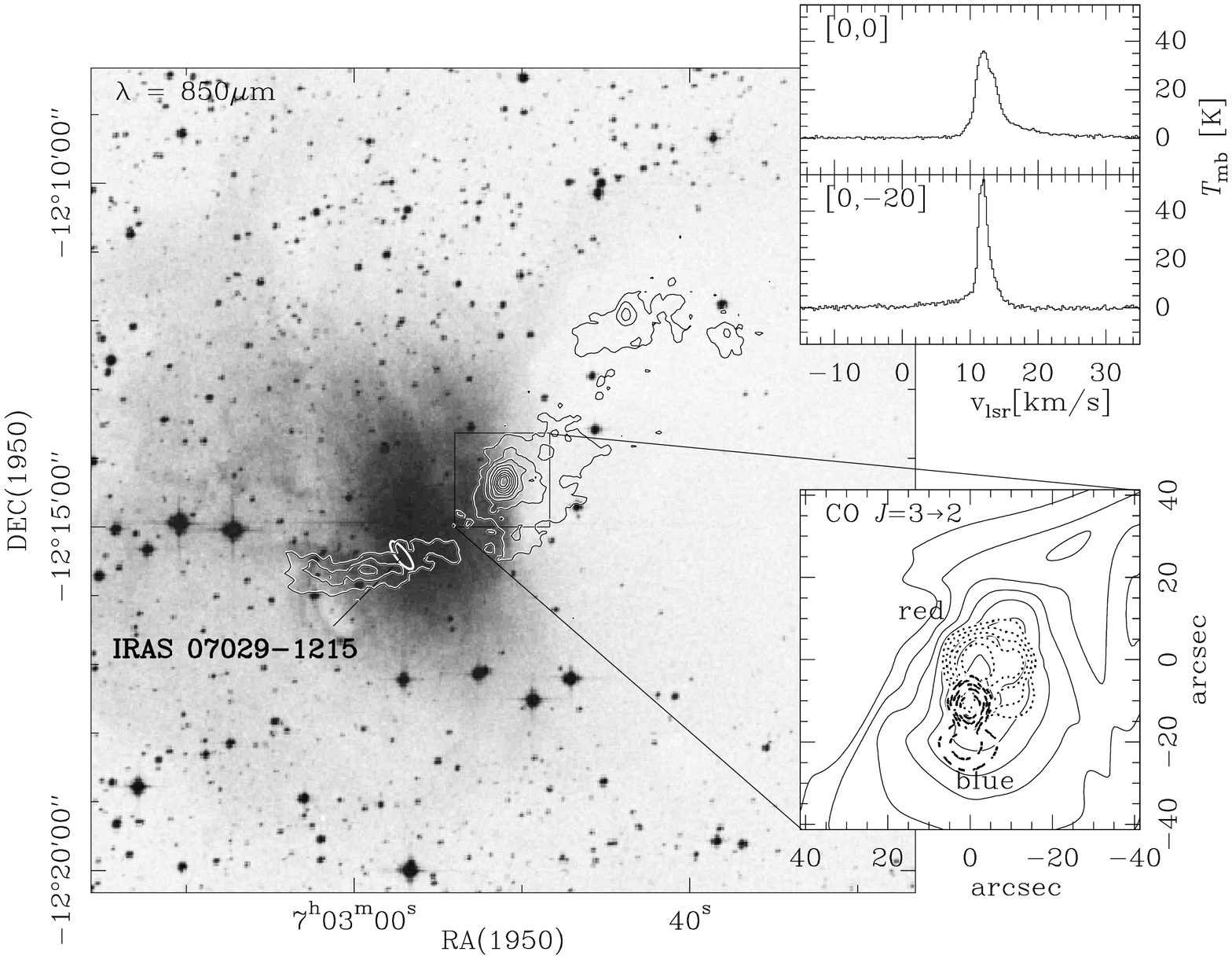}}} 
   \end{minipage}
   \hspace{1.0cm}
   \begin{minipage}[t]{10.0cm}
   \vspace{0.5cm}
  Figure 1. Left: UYSO 1, seen in $\lambda=850$ $\mu$m continuum emission (contours, background: POSS). Upper Right: Two examples of the CO($J=3 \to 2$) line wings. Lower Right: Velocity-integrated emission of the CO line and its wings.
   \end{minipage}
   \end{figure}
%
%

\section{Introduction}
\label{sec:1}
Whether the formation of massive stars significantly differs from the formation of low mass stars is not yet fully understood. While disk accretion seems possible under special circumstances (e.g. \cite{jia96}), coalescence of smaller objects into one massive star is an alternative model (e.g. \cite{bon98}). Observations of molecular outflows in regions of massive star formation indicate, however, that the accretion model is indeed applicable.
Even though little is known about the initial conditions of \index{massive star formation} \cite{eva02}, young massive protostars are expected not to be detectable at NIR/MIR wavelengths, but mainly in the submm continuum. 
\\ \\
Using SCUBA and IRAM bolometers, we investigated a sample of 47 luminous IRAS sources, searching for massive protostellar objects in their close vicinities \cite{bpda, kle0X}. Near IRAS 07029-1215, we discovered a deeply embedded object in a distance of about 1 kpc \cite{wbr89}, powering a high-velocity bipolar CO outflow. The IRAS source is located in the bright H\,{\sc ii} region S 297, it has an IRAS luminosity of 1700 L$_{\odot}$ \cite{hen92}.

\section{Observations}
\label{sec:2}
Directly following the SCUBA continuum observations at $\lambda = 450$ $\mu$m and $\lambda = 850$ $\mu$m taken in October 1999, a map in CO($J=3 \to 2$) was obtained using the B3 facility receiver). Additional observations in CS($J=2 \to 1$), CS($J=5 \to 4$), SiO($J=2 \to 1$) and especially H$_2$CO($J=3_{03}-2_{02}$)/($J=3_{22}-2_{21}$) were carried out at the IRAM 30m telescope in March 2003.

\section{Results}
\label{sec:3}
The surroundings of IRAS 07029-1215 are shown in Fig.~1 as a POSS image. The SCUBA 850 $\mu$m data, overlaid as contours, show UYSO 1 as a compact emission peak at $\alpha_{\rm (B1950.0)}=07^{\rm h}02^{\rm m}51^{\rm s} , \delta_{\rm (B1950.0)}=-12^{\circ}14'26"$, between S~297
and a dark cloud. Two additional dusty filaments are visible at 850 $\mu$m, one of them containing the IRAS source. 
\\ \\
UYSO 1 was not detected at near- and mid-infrared wavelengths ($\lambda = $2.2.. 20~$\mu$m) by MSX and 2MASS.
\\ \\
The CO map in Fig.~1 (lower right) is a close-up view of UYSO 1. The line wings extending from -20 km s$^{-1}$ at [0",-20"] to +40 km s$^{-1}$ at [0",0"] can be clearly distinguished from well-determined baselines. These line wings are due to a high-velocity bipolar outflow originating from UYSO 1. The outflow is less clearly visible in the CS data, however lines on respective positions have corresponding shapes. The spatial resolution of the observations as of now is insufficient to conclude whether the bipolar outflow is composed of only one or more components \cite{beu02}.

\section{Analysis}
\label{sec:4}
\subsection{Temperature and Mass Estimates}
A temperature estimate was obtained from the H$_2$CO($J=3_{03} \to 2_{02}$)/($J=3_{22} \to 2_{21}$) line ratio, following \cite{man93}, resulting in T=40$\pm 5$ K for a relatively wide density range.
\\ \\
Following \cite{hen00}, we derive hydrogen column densities from the CO $J=3\to 2$ transition. This leads to $M_{\rm gas}$ = 40 M$_{\odot}$. Another estimate for the total mass has been calculated by \cite{bpda} from an analysis of the continuum emission at $\lambda = 850$ $\mu$m, resulting in $M=15$ M$_{\odot}$. The discrepancy can at least partly be explained by a smaller region taken into account in the latter case as well as the dust continuum emission tracing only the high-density central regions while the CO emission is collected along the line of sight throughout the low-density halo. Thus, the envelope mass estimates based on line and dust emission are compatible when keeping in mind their respective limitations and range between 15 and 40 M$_{\odot}$. The virial mass is much higher ($M_{\rm vir}$ = 180 M$_{\odot}$), which is certainly caused by an overestimation of line width due to optical thickness as well as turbulent motion, in addition to the cloud not being in virial equilibrium.

\subsection{Outflow Properties}
The CO line wings were analyzed to their respective 10\% contour levels, corresponding to a 2 $\sigma$ detection for the redshifted line wing and a 4 $\sigma$ detection for the blueshifted line wing. The masses contained inside the two outflow lobes were calculated by integration of H$_2$ column densities, resulting in $M_{\rm blue}$ = 1.8 M$_{\odot}$ and $M_{\rm red}$ = 3.6 M$_{\odot}$, the total outflow mass thus being $M_{\rm out}$ = 5.4 M$_{\odot}$.
An analysis of the dynamical timescale, given for an inclination of $i=57.3^{\circ}$ (the most probable value, cf. \cite{bon96}), leads to  $t_d(i=57.3^{\circ})=1700$ yrs. We note that this is not a  good age indicator. The corresponding mass outflow rate is $\dot M(57.3^{\circ}) = 3.3 \times 10^{-3}$ M$_{\odot}$/yr. Only a massive central object is able to uphold such an outflow rate.

\subsection{Spectral Energy Distribution}

Given the measured continuum fluxes at 450 $\mu$m and 850 $\mu$m, flux estimates from the IRAS atlas images and NIR/MIR upper limits from MSX and 2MASS, a tentative SED of UYSO 1 could be derived. Integration leads to a luminosity estimate of $<1900$ L$_{\odot}$. A modified blackbody accounting for the dust optical depth was fitted to the data with $T=45$ K and $\beta=2$. The continuum emission was then modeled using the radiative transfer code developed by \cite{man98} with standard ISM parameters and in 1D mode. A model with an envelope mass of $M_{\rm env}$ = 30 M$_{\odot}$ and an assumed single central object in the B2.5 class fits the data well.

\section{Conclusions}

The emerging picture of UYSO 1 is that of an early B star surrounded by an 
envelope of about 30-40 M$_{\odot}$. The object seems to be in a particularly early evolutionary state, however already shows a bipolar outflow with a total mass of M$_{\rm outflow}$ = 5.4 M$_{\odot}$. To our knowledge, there are only very few comparable sources that are still undetected at infrared wavelengths (see e.g. \cite{fon03}). With additional high-resolution observations, we hope to learn more about this early evolutionary stage of massive star formation. For more details, see \cite{for03}.

%
%
%
%
%
%
%
%
%

%
%



\printindex
\end{document}